\documentclass[showpacs,twocolumn,nofootinbib,pra]{revtex4}
\usepackage{doi}%
\usepackage{hyperref}

\usepackage[dvips]{epsfig}
\usepackage{amssymb}
\usepackage{amsmath}
\usepackage{latexsym}
\usepackage{epsfig}
\usepackage{graphicx}
\usepackage{color}
\usepackage{fouridx}
\newcommand{\mbfp}{\mathbf{p}}
\newcommand{\hU}{\hat{U}}
\newcommand{\hD}{\hat{D}}
\newcommand{\hH}{\hat{H}}
\newcommand{\hadag}{\hat{a}^{\dagger}}
\newcommand{\ha}{\hat{a}}
\newcommand{\clA}{\mathcal{A}}

\newcommand{\clT}{{\cal T}}
\newcommand{\mIm}{\mbox{Im}\,}
\newcommand{\rgl}{\rangle}
\newcommand{\lgl}{\langle}

\newcommand{\rAlpha}{{|\alpha\rangle}}
\newcommand{\lAlpha}{{\langle\alpha |}}
\newcommand{\clK}{{\cal K}}

\newcommand{\hclH}{\hat{\cal H}}

\newcommand{\clH}{{\cal H}}

\newcommand{\hatp}{\hat{p}}
\newcommand{\hatz}{\hat{z}}

\newcommand{\prt}{\partial}

\newcommand{\mbfz}{\mathbf{z}}
\newcommand{\hbdag}{\hat{b}^{\dagger}}
\newcommand{\hb}{\hat{b}}

\newcommand{\ep}{\epsilon}

\begin{document}

\title{On quantum free-electron laser: Superradience}

\author{Alexander Iomin}

\affiliation{Department of Physics, Technion, Haifa, 32000,
Israel}


\begin{abstract}
A quantum model of a free-electron laser is considered 
for the many electron system. An exact expression for 
the evolution of the laser amplitude is obtained in the framework of
the coherent state consideration. 
Reliable conditions for the superradiance of the high-gained laser
is discussed for the short time limit of the exact solution. 

\end{abstract}

\pacs{2.50.-p, 42.55.-f, 3.65.-w}

\maketitle

\section{Introduction} 
Experimental implementation and theoretical description of 
free-electron lasers is a long-lasting problem that started 
in seventies of the last century. This extensively studied 
phenomenon is well described and reviewed 
\cite{BonEtAl90,colson90,DaReTo93,SaScYu2000}, 
to mention a few. Contemporary studies are also reflected 
in recent publications and related to both classical and 
quantum tasks \cite{PiVo2021,KGCSS21}, including application of the fractional calculus to the FEL model 
\cite{ArDaLiPa2017,iom21}.

In this notes we consider a quantum model of a free-electron laser (FEL),
which has been obtained from the consideration of a 
non-relativistic electron in an electromagnetic field 
\cite{BeMc83,BeMc87} in a so-called Bambini–Renieri frame 
\cite{BaRe78,BaReSt79}, see also App. \ref{q-fel-H}. For a system of $N_e=N$ electrons, 
the quantum Hamiltonian reads \cite{BeMc83,BeMc87,KGEPS15,KGCSS19}
\begin{equation}\label{fel-1}
\hH
=\sum_{j=1}^{N}\frac{\hatp_j^2}{2m}+\hbar g\left(
\ha\sum_{j=1}^{N}e^{2ik\hatz_j}+\hadag\sum_{j=1}^{N}e^{-2ik\hatz_j}
\right),
\end{equation}
where summation relates to the positions $\hatz_j$ and momenta 
$\hatp_j$ of electrons with the mass $m$ and the wave number $k$.
The position-momentum commutation rule is 
$[\hatz_k,\hatp_j]=i\hbar\delta_{k,j}$.
The laser mode is described by the photon annihilation and 
creation operators $\ha$ and $\hadag$, respectively 
with the commutation rule
$[\ha,\hadag]=1$. The interaction parameter $g$ couples 
the electron dynamics to the photon laser field.

Various approaches have been considered and discussed in the 
literature \cite{BoPiRo05,KGEPS15,KGCSS19}, 
see an extended discussion in recent results \cite{KGCSS19,KGCSS21}.
The problem is well posed and discussed in the literature,
and our main concern in this paper relates to the technicality 
of the exact solution and its short time approximation.

Our main question in task is the intensity of the laser 
field amplitude 
\begin{equation}\label{fel-2}
\clA(t)=\langle\hadag(t)\ha(t)\rangle=\langle\Psi_0|
\hU^{\dagger}(t)\hadag\ha\hU(t)|\Psi_0\rangle  ,
\end{equation}
where $\hU(t)$ is the unitary evolution operator. The initial wave
function is the direct product of the photon and electron wave functions
\begin{equation}\label{fel-3a}
|\Psi_0\rangle = |\alpha\rangle\otimes |\mathcal{P}\rangle ,
\end{equation}
where the coherent states \cite{glauber} $|\alpha\rangle$ are chosen for the photon wave function \cite{BeScZu82}, 
while the electron wave function can be different realization of 
the electron initial conditions. In particular it can be either an 
orthogonal basis
$|\mathcal{P}\rangle =|\mbfp\rangle=\prod_{j=1}^{N}|p_j\rangle\equiv |p_1,\dots,p_N\rgl$, where
$\langle p_k|{p'}_j\rangle=\delta_{k,j}$, or the electron coherent 
states $|\mathcal{P}\rangle =|\vec{\beta}\rangle$. We consider the both.

It is worth be stressing that discussing the dynamics of the 
system we also set the notation for the operators considered here. 
Namely, for any time dependent operator
$\hat{x}(t)\equiv\hat{x}_t $ its initial value is denoted by 
$\hat{x} =\hat{x}(t=0)$. The same is concerned with its average 
values denoting $x(t)\equiv x_t$ and $x=x(t=0)$, where 
$x= (a^*,a, \{p_j,z_j\})$.
The only exclusion is for the Hamiltonian, since it is 
the integral of motion and correspondingly $\hclH(t)=\hclH$.

We start from the Heisenberg equation of motion for the dynamics of the laser field intensity in the operator form
$\hat{\clA}(t)=\hadag(t)\ha(t)$,
\begin{equation}\label{int2}
\dot{\hat{\clA}}(t)=\frac{1}{i\hbar}[\hat{\clA}(t), \hclH] .
\end{equation}
In this approach, the evolution of the 
averaged value of the laser field intensity is defined as follows
$$\langle\hat{\clA}(t)\rangle =
e^{\clK t}\langle\hat{\clA}(t=0)\rangle ,$$
where $\clK$ is a so-called Koopman operator \cite{koopman},
see also \textit{e.g.}, \cite{gaspard,cvitanovic}.

Another approach, developed in the paper is investigation 
of the evolution of the initial wave function, 
which relates to the construction of the evolution 
operator by means of the path integrals \cite{feynman,schulman}.
For the latter example, in the initial wave functions 
\eqref{fel-3a}, the electron wave function is the
direct product of electron wave functions
\begin{equation}\label{fel-3}
 |\mathcal{P}\rangle =  |\mbfz\rangle =
 \prod_{j=1}^{N}|z_j\rangle\equiv |z_1,\dots,z_N\rgl
\end{equation}
where 
$\langle z_k|{z'}_j\rangle=\delta_{k,j}\delta(z-z')$.

Another possibility is the momentum representation of the electron
wave function, when $\hatp_j|\mbfp\rangle=p_j|\mbfp\rangle $ and
$\hatz_j=i\hbar\frac{d}{dp_j}$, 
the operator $e^{-2ik\hatz_j}$ acts as a shift operator
\begin{equation}\label{fel-4}
e^{-2ik\hatz_j}|p_j\rangle=e^{2k\hbar\frac{d}{dp_j}}|p_j\rangle=
|p_j+2k\hbar \rangle.
\end{equation}
Analogously, $\langle p_j+2k\hbar |=\langle p_j|e^{2ik\hatz_j} =
\left(e^{-2ik\hatz_j}|p_j\rangle\right)^{\dagger}$.

\section{Koopman operator}

For the quantum mechanical analysis of the Heisenberg equations
of motion 
we use a technique of mapping the Heisenberg equations on a basis
of the coherent states \cite{st1982,biz1981,bikt1986, iom16,iom17}. 
Since the Hamiltonian is the integral of motion
then it is time independent, $\hclH(t)=\hclH(t=0)$. 
Therefore, the Hamiltonian in the Heisenberg equations
can be mapped on a basis of the coherent states 
$|\alpha\rgl$ and $|\vec{\beta}\rgl$, constructed 
at the initial moment $t=0$..
First, we construct the basis of the photon coherent states.
That is, at the initial moment $t=0$, one introduces 
the coherent states vector $\rAlpha$ as the eigenfunction
of the annihilation operators $\ha=\ha(t=0)$, such that 
$\ha(t=0)|\alpha\rgl\equiv \ha|\alpha\rgl=\alpha|\alpha\rgl $
and correspondingly 
$\lgl\alpha |\hadag =[\ha|\alpha\rgl]^{\dag}=\alpha^*\lgl\alpha |$.
The coherent state can be also constructed from a vacuum state 
$|0\rgl$ as follows \cite{perelomov}
\begin{equation}\label{koop-1}
\rAlpha=\exp[\alpha\hadag-\alpha^{*}\ha]|0\rgl, \quad
\ha|0\rgl=0 .
\end{equation}
Then averaging the photon field operator over the photon coherent states $\lAlpha\hat{\clA}(t)\rAlpha$, 
one maps the Heisenberg equation of motion \eqref{int2}
on the basis of the coherent states $\rAlpha$
as follows
\begin{equation}\label{koop-3}
i\hbar\frac{d}{dt}\lAlpha\hat{\clA}(t)\rAlpha
=\langle\alpha|
\left[\hat{\clA}\hat{\clH}-
\hat{\clH}\hat{\clA}(t)\right]|\alpha\rangle .
\end{equation}
Taking into account Eq. \eqref{koop-1}, one obtains 
the mapping rules \cite{KlSu68}
\begin{subequations}\label{koop-4}
\begin{align}  
&\lAlpha\hat{\clA}(t)\hadag\rAlpha =
e^{-|\alpha|^2}\frac{\prt}{\prt\alpha}e^{|\alpha|^2}
\clA(t) , \label{koop-4a} \\
&\lAlpha\ha\hat{\clA}(t)\rAlpha =
e^{-|\alpha|^2}\frac{\prt}{\prt\alpha^{*}}e^{|\alpha|^2}\clA(t) .
\label{koop-4b} 
\end{align}
\end{subequations}

The next step is the averaging procedure for the 
electron part of the Hilbert space. 
To this end we construct a basis of the electron coherent 
states at the initial moment $t=0$. 
We admit that the electron system in the Hamiltonian \eqref{fel-1}
is considered as a system of free spinless 
particles and their commutation 
rules corresponds to a so called Heisenberg - Weyl group
\cite{perelomov}. Therefore introducing creation 
$\hbdag$ and annihilation $\hb$ operators:
\begin{subequations}\label{koop-8}  
\begin{align}
&\hb_j=\frac{z_j+i\hatp_j}{\sqrt{2\hbar }},  \quad \quad
\hbdag=\frac{z_j-i\hatp_j}{\sqrt{2\hbar }},  \label{koop-8a} \\
& z_j=\sqrt{\frac{\hbar}{2}}(\hb_j+\hbdag_j),   \quad
\hatp_j=i\sqrt{\frac{\hbar}{2}}(\hbdag_j-\hb_j),   \label{koop-8b}
\end{align}
\end{subequations}
where $[\hb_j,\hbdag_{j'}]=\delta_{j,j'}$,
we introduce the coherent state basis 
\begin{equation}\label{koop-9}
|\vec{\beta}\rgl=\prod_{j=1}^N|\beta_j\rgl =
\prod_{j=1}^N\exp[\beta_j\hbdag_j-\beta_j^{*}\hb_j]|0\rgl, \quad
\hb_j|0\rgl=0 .
\end{equation}
which belongs to the electron Hilbert space.

Introducing the double average 
\begin{equation}\label{koop-10}
\clA(t)\equiv\lgl\hat{\clA}(t)\rgl= 
\lgl\lgl\hat{\clA}(t){\rgl}_{\alpha}{\rgl}_{\beta}
\end{equation}
and using properties \eqref{koop-4} for both 
$\ha,\hadag$ and $\hb,\hbdag$ operators and taking into account 
the commutator $[\hat{\clA},\mathbf{p}^2]$, we obtain the ``Koopman equation''
\begin{equation}\label{koop-11}
\frac{d}{dt}\clA(t)=\clK\clA(t),
\end{equation}
where the Hermitian Koopman operator reads
\begin{subequations}\label{koop-12}
\begin{align}
\clK= & \frac{i}{4m}\sum_{j=1}^N\left(\partial_{\beta^*_j}^2
-\partial_{\beta_j}^2+2\beta_j\partial_{\beta^*_j} \right.
\nonumber \\
- & \left.
2\beta_j^*\partial_{\beta_j} +2\beta_j\partial_{\beta_j}-
2\beta_j^*\partial_{\beta^*_j}\right)  \label{koop-12a} \\
 & +ig\sum_{j=1}^N 
 \left[\partial_{\alpha^*}e^{2ik(\beta_j^*+\beta_j)}
e^{2ik\partial_{\beta_j^*}}\right. \nonumber \\
- & \left.
\partial_{\alpha}
e^{-2ik(\beta_j^*+\beta_j)}e^{-2ik\partial_{\beta_j}}\right] .
\label{koop-12b}
\end{align}
\end{subequations}
The Koopman operator consists of two parts: 
$\clK=\clK_e+\clK_{e-ph}$. The first part $\clK_e$, described 
by Eq. \eqref{koop-12a} corresponds to the kinetic part of
free electrons, while the second part describes 
the electron-photon interaction, Eq. \eqref{koop-12b}.
The second part is more sophisticated operator and its inferring 
needs some care see App. \ref{koop-operator}.

\subsection{Exact solutions}\label{ex-sol-I}

An exact expression can be obtained from the formal solution
\eqref{koop-13}. Let us consider the Koopman equation
\eqref{koop-11},
\begin{equation}\label{ex-sol-I-1}
\frac{d}{dt}\clA(t)=\clK\clA(t)
\end{equation}
with the solution 
\begin{equation}\label{ex-sol-I-2}
\clA(t)=e^{t\clK} \alpha^*\alpha ,
\end{equation}
and let the Koopman operator is such that
\begin{equation}\label{ex-sol-I-3}
\clK\clA(t=0)=\clK\alpha^*\alpha =e \alpha^*\alpha,
\end{equation}
where  is a constant value. This stationary equation leads to the exponential gain solution 
$$\clA(t)=e^{et}|\alpha|^2$$ 
with $e>0$.

Another important and exact result in the electron coherent states
is the constant values of the averaged energy and the moment of electrons. This conservation law immediately follows from 
the specific form of the Koopman operator \eqref{koop-12} and the transformations
\eqref{koop-8}.
Namely $$\clK \langle p_j^2\rangle=0 ~~ \text{and}~~ 
\clK \langle p_j\rangle=0, $$ 
then 
$\langle p_j^2(t)\rangle=\langle p_j^2\rangle$ and
$\langle p_j(t)\rangle=\langle p_j\rangle$. This situation 
is a pure quantum effect. It
can be explained by coherent Cherenkov radiation, which is described just by the same Hamiltonian as in Eq. \eqref{fel-1} \cite{BeMc87},
see also Appendix \ref{q-fel-H}.

\subsection{Short time approximation I}\label{sta-I}

The solution of the Koopman equation \eqref{koop-11}
for the initial condition  $\lgl\clA(t=0)\rgl=\alpha^*\alpha$ 
is presented in the exponential form \cite{iom2022}
\begin{equation}\label{koop-13}
\clA(t)=e^{t\clK} \alpha^*\alpha
=\sum_{n=0}^{\infty}\frac{t^n}{n!}\clK^n\alpha^*\alpha .
\end{equation}
We take into account that electrons spend a very short time 
inside the laser size, namely $t\ll 1$, 
which results from the fact that the laser size is 
finite and the electron velocity is very large (even large than the speed of light). Then we should 
take into account only few first terms in the expansion 
\eqref{koop-13}, and we restrict ourselves by the second 
order of  $t^2$.
Obviously, for $t^0$, the zero order term corresponds 
to the initial condition: ${\lgl\clA\rgl}^{(0)}=|\alpha|^2$. 
Correspondingly, the first order term for $t^1$ reads
\begin{multline}\label{koop-14}
{\lgl\clA\rgl}^{(1)} =t\clK\alpha^*\alpha=
igt\sum_{j=1}^N \left[\partial_{\alpha^*}e^{2ik(\beta_j^*+\beta_j)}
e^{2ik\partial_{\beta_j^*}}- \right. \\
\left. \partial_{\alpha}
e^{-2ik(\beta_j^*+\beta_j)}e^{-2ik\partial_{\beta_j}}\right]
\alpha^*\alpha \\
=igt\sum_{j=1}^N \left[\alpha e^{2ik(\beta_j^*+\beta_j)}
-\alpha^*e^{-2ik(\beta_j^*+\beta_j)}\right].
\end{multline}
At this step, the action of the Koopman operator reduces 
to differentiation 
with respect to $\alpha$ and $\alpha^*$ only. The derivatives 
$\partial_{\beta_j}$ and $\partial_{\beta_j^*}$ do not ``work''
here. The situation changes for the second order term
for which we obtain
\begin{multline}\label{koop-15}
{\lgl\clA\rgl}^{(2)} =\frac{t}{2}\clK {\lgl\clA\rgl}^{(1)} =
\clK_e{\lgl\clA\rgl}^{(1)} -\frac{g^2t^2}{2} \\
\times\sum_{j=1}^N \left[\partial_{\alpha^*}e^{2ik(\beta_j^*+
\beta_j)}
e^{2ik\partial_{\beta_j^*}}-\partial_{\alpha}
e^{-2ik(\beta_j^*+\beta_j)}e^{-2ik\partial_{\beta_j}}\right] \\
\times
\sum_{j=1}^N \left[\alpha e^{2ik(\beta_j^*+\beta_j)}
-\alpha^*e^{-2ik(\beta_j^*+\beta_j)}\right] \\
= 
\frac{t}{2}\clK_e {\lgl\clA\rgl}^{(1)} 
 +e^{4k^2}\frac{g^2t^2}{2} \sum_{j',j=1}^N
e^{2ik(\beta_j^*+\beta_j-\beta_{j'}^*+\beta_{j'})}.
\end{multline}

We restrict the expansion by the second order in Eq. 
\eqref{koop-15}. The last term is the main 
contribution 
to laser amplitude 
\begin{equation}\label{koop-16}
\clA(t)\approx
e^{4k^2}\frac{g^2t^2}{2} \sum_{j',j=1}^N
e^{2ik(\beta_j^*+\beta_j-\beta_{j'}^*+\beta_{j'})},
\end{equation}
and it is the first main result, which will be discussed 
in Sec. \ref{disc}

\section{Interaction representation} 

Let us consider the evolution operator in the interaction representation 
with the Hamiltonian\footnote{It is worth noting that the Hamiltonian \eqref{fel-1} is already presented in the interaction representation with respect to a photon term 
$\hbar\omega\hadag\ha$, where $\omega$ 
is the photon frequency \cite{BeMc83}. 
However the action of this operator reduces to a simple
time shift of the initial phase of the coherent state 
$|\alpha\rgl\rightarrow |e^{-i\omega t}\alpha\rgl$.
Note also that in the Bambini-Renieri frame, $\omega=kc$ \cite{KGEPS15}, where $c$ is the light speed.}
$\hH_I=\hH_I(t)$, which can be obtained 
by the chain of transformations as follows
\begin{multline}\label{fel-5}
\hH_I=\hbar g e^{i\frac{\hat{\mbfp}^2t}{2m\hbar}}
\left(
\ha\sum_{j=1}^{N}e^{2ik\hatz_j}+\hadag\sum_{j=1}^{N}e^{-2ik\hatz_j}
\right) e^{-i\frac{\hat{\mbfp}^2t}{2m\hbar}}  \\
=\hbar g \sum_{j=1}^{N}\left[\ha
\exp\left(2ik e^{i\frac{\hat{\mbfp}^2t}{2m\hbar}}\hatz_j
e^{-i\frac{\hat{\mbfp}^2t}{2m\hbar}}\right)\right. + \\
+\left.\hadag\exp\left(-2ik e^{i\frac{\hat{\mbfp}^2t}{2m\hbar}}
\hatz_j e^{-i\frac{\hat{\mbfp}^2t}{2m\hbar}}\right)\right] = \\
=\hbar g\sum_{j=1}^{N}\left[\ha e^{2ik(\hatz_j+t\hatp_j/m)}
+\hadag e^{-2ik(\hatz_j+t\hatp_j/m)}\right] .
\end{multline}
Here we use that 
$
e^{i\hat{\mbfp}^2}f(\hatz_j)e^{-i\hat{\mbfp}^2}=
f\left(e^{i\hat{\mbfp}^2}\hatz_j e^{-i\hat{\mbfp}^2}\right) 
$
and the commutation rule \eqref{fel-4} in the momentum 
representation of the electron wave function,
\begin{equation}\label{fel-4a}
i\hbar\left[\frac{d}{d p_j} , e^{-\frac{it}{2m\hbar}p_j^2}\right]=
\frac{t}{m}e^{-\frac{it}{2m\hbar}p_j^2}p_j .
\end{equation}
Correspondingly, the evolution operator in the interaction 
representation according to Eq. \eqref{fel-5} reads
\begin{equation}\label{fel-6}
\hU_I(t)=\hat{\clT}\exp\left[-\frac{i}{\hbar}\int_0^t\hH_I(\tau)d\tau 
\right] ,
\end{equation}
where $\hat{\clT}$ is the time ordering operator.

The next step of the treatment of Eq. \eqref{fel-6} 
is the standard procedure of the partition of the time 
interval $t=R\Delta t$ as for the path integral 
construction at the condition $\Delta t\rightarrow 0$, while
$R\rightarrow\infty$. Taking $\tau_r=rt/R$ as the 
center of the $r$-th $\Delta t$-segment, we obtain
\begin{multline}\label{fel-7}
\hU_I(t) =\lim_{R\to\infty} e^{-\frac{i}{\hbar}\hH_I(\tau_R)\Delta t}
\dots \\
\dots
e^{-\frac{i}{\hbar}\hH_I(\tau_r)\Delta t}\dots 
e^{-\frac{i}{\hbar}\hH_I(\tau_1)\Delta t} .
\end{multline}
Every $r$-th exponential has the form of the following unitary operator
 $\hD(\alpha)=\exp(\alpha\hadag-\alpha^*\ha)$, where
$\hD\hD^{\dagger}=\hD^{\dagger}\hD=1$, \cite{CaNi68}. From Eqs. 
\eqref{fel-5}, \eqref{fel-6}, and \eqref{fel-7}, we have
\begin{subequations}\label{fel-8}
\begin{align}
\hD(\alpha_r)= &\exp\left[\alpha_r(\tau_r)\hadag -
\alpha_r^*(\tau_r)\ha\right], 
\label{fel-8a} \\
\alpha_r(\tau_r)= & -ig\Delta t\sum_{j=1}^{N}
 e^{-2ik(\hatz_j+\tau_r\hatp_j/m)} , \label{fel-8b} \\
\alpha_r^*(\tau_r)= & \, ig\Delta t\sum_{j=1}^{N}
 e^{2ik(\hatz_j+\tau_r\hatp_j/m)} . \label{fel-8c}
\end{align}
\end{subequations}
Here $\alpha_r$ and $\alpha_r^*$ for $r=1,\dots, R$ 
are the operator valued functions of the electron momentum and coordinate $p_j$ and $\hat{z}_j$ respectively.

In the interaction representation, the laser field amplitude
\eqref{fel-2} reads
\begin{equation}\label{fel-9}
\clA(t)={}_I\langle\Psi_0|
\hU_I^{\dagger}(t)\hadag\ha\hU_I(t)|\Psi_0{\rangle}_I  ,
\end{equation}
while the initial wave function \eqref{fel-3} is transformed
as follows
\begin{equation}\label{fel-10}
|\Psi_0{\rangle}_I = |\alpha\rangle\otimes 
e^{-i\frac{\mbfp^2t}{2m\hbar}}|\mbfp\rangle .
\end{equation}

\subsection{Short time approximation II}

The first step of the averaging procedure is evaluation of 
the action of the evolution operator on the coherent states 
$\hU_I(t)|\alpha\rangle$ that eventually reduces to the 
action of the shift operators $\hD(\alpha_r)$.
This procedure is well defined \cite{perelomov}, and the operators
$\hD(\alpha_r)$ transform any coherent state to another coherent state,
\begin{equation}\label{fel-11}
\hD(\alpha_r)|\alpha\rangle=e^{i\mIm (\alpha_r\alpha^*)}
|\alpha+\alpha_r\rangle .
\end{equation}
Therefore, following this procedure, we obtain 
\begin{equation}\label{fel-12}
|\alpha(t)\rgl =\prod\limits_{r=1}^R\hD[\alpha_r(\tau_r)]|\alpha\rangle 
\end{equation}

Then from Eqs. \eqref{fel-9} and \eqref{fel-12} we obtain
\begin{multline}\label{fel-14}
\lgl\alpha(t)|\hadag\ha|\alpha(t)\rgl \\
=
\lgl\alpha|\left(\prod\limits_{r=1}^R\hD[\alpha_r(\tau_r)]\right)^{\dagger}
\hadag\ha\prod\limits_{r=1}^R\hD[\alpha_r(\tau_r)]|\alpha\rgl .
\end{multline}

An important feature of 
the analysis is the operator valued complex functions 
$\alpha_r$, which ``do not work'' as $c$-numbers, since they 
are not commute. However, in the short time dynamics, considered in 
Sec. \ref{sta-I}, we have $\tau_r\sim \Delta t\sim t$ and correspondingly $r=R=1$. Therefore, on the short time scale 
Eq. \eqref{fel-14} is simplified and reads
\begin{multline}\label{fel-21}
\lgl\alpha(t)|\hadag\ha|\alpha(t)\rgl 
\approx
\lgl\alpha|\left(\hD[\alpha_1(t)]\right)^{\dagger}
\hadag\ha\hD[\alpha_1(t)]|\alpha\rgl  \\
= \left[\alpha^*+\alpha_1(t)\right]\cdot
\left[\alpha +\alpha_1(t)\right] +|\alpha|^2 \\
= -ig\alpha^* t
\sum_{j=1}^{N} e^{-2ik(\hatz_j+t\hatp_j/m)}
+ig\alpha t\sum_{j=1}^{N} e^{2ik(\hatz_j+t\hatp_j/m)} \\
+g^2t^2
\sum_{j=1}^{N} e^{2ik(\hatz_j+t\hatp_j/m)} \cdot
\sum_{j'=1}^{N} e^{-2ik(\hatz_{j'}+t\hatp_{j'}/m)}.
\end{multline}

\subsection{Electron wave functions}
The obtained result in Eq. \eqref{fel-21} 
is a self-adjoined operator in the Hilbert space
of the electron wave functions
$e^{-i\frac{\mbfp^2}{2\hbar m}t}|\mbfp\rgl$.
According to the Baker-Hausdorff theorem the operator valued exponential functions 
in Eq. \eqref{fel-21} can be present as follows \cite{louisell},
\begin{equation}\label{fel-22}
e^{- 2ik(\hatz_j +\tau\hatp_j/m)} =e^{2i\frac{\hbar k^2}{m}\tau}
e^{- 2ik\tau\frac{\hatp_j}{m}}e^{2k\hbar\frac{d}{dp_j}}
\end{equation}
for the $\alpha_1(t)$ operator valued function and the Hermitian conjugate form 
for the $\alpha^{\dagger}_1(t)$.

Therefore, the laser field amplitude in Eq. \eqref{fel-9} reduces 
to the averaging procedure over the electron wave functions
\begin{multline}\label{fel-23}
\clA(t)= \\
\langle \mbfp|e^{i\frac{\mbfp^2t}{2m\hbar}}
\left[\alpha^*+ 
ig t\sum_{j=1}^{N}e^{- 2k\hbar\frac{d}{dp_j}}
e^{-i2\frac{\hbar k^2}{m}t}\, e^{2ikt\frac{\hatp_j}{m}}
\right] \times \\
 \left[\alpha -igt\sum_{j=1}^{N} e^{2i\frac{\hbar k^2}{m}t}
e^{-2ikt\frac{\hatp_j}{m}} \,e^{2k\hbar\frac{d}{dp_j}}
\right]
e^{-i\frac{\mbfp^2t}{2m\hbar}}|\mbfp\rangle .
\end{multline}
In the averaging procedure, let us first estimate
only the diagonal matrix elements, which remain,
due to the shift operations \eqref{fel-4} 
$e^{\pm 2k\hbar\frac{d}{dp_j}}|\mbfp\rgl
=|p_1,\dots, p_{j-1}, p_j\pm 2k\hbar, p_{j+1}\dots\rgl$.
Therefore, due to the orthogonality of the wave functions 
this yields
\begin{multline}\label{fel-24}
\Big\langle \mbfp\, e^{i\frac{\mbfp^2t}{2m\hbar}}\,
e^{- 2k\hbar\frac{d}{dp_j}}
\Big| e^{-2ik(t-t)\frac{\hatp_j}{m}}  \Big|
 e^{2k\hbar\frac{d}{dp_j}} e^{-i\frac{\mbfp^2t}{2m\hbar}}\,\mbfp
 \Big\rangle =1 \\
 {}
\end{multline}
and Eq. \eqref{fel-23} reduces to the following expression
\begin{equation}\label{fel-25}
\clA(t)=|\alpha|^2+g^2t^2N.
\end{equation}

\section{Discussion: Superradiance}\label{disc}

We collect some physical consequences of the calculations
of the laser intensity and shall discuss them in this section.
The main contribution to the intensity of the laser photon field 
is due to the interaction term in the Hamiltonian 
\eqref{fel-1}, that is
\begin{equation}\label{disc-1}
\clA(t)\sim g^2\left\lgl\left[
\ha\sum_{j=1}^{N}e^{2ikz_j}+\hadag\sum_{j=1}^{N}e^{-2ikz_j}
\right]^2\right\rgl .
\end{equation}
Two different approaches for the evaluation of the laser intensity
are suggested in the framework of the Heisenberg and 
Schr\"odinger representations of quantum mechanics. 

In the first approach, the Heisenberg equations of motion
are mapped on the basis of the coherent states, which are 
constructed at the initial moment of time $t=0$ for both the 
photon laser field and electrons. Since the algebras of 
the photon and electron operators belong to the same 
Heisenberg-Weyl algebras,
the analytical forms of the coherent states 
(as the eigenstates of the annihilation operators) 
are the same, although the wave 
functions $\rAlpha$ and $|\vec{\beta}\rgl$
belong to the different Hilbert spaces.
The Heisenberg equations of motions for the laser
photon field intensity
are exactly mapped on the basis
of the coherent states and the obtained ``Koopman equation''
\eqref{koop-11}
describes the gain process and it is controlled 
by the Koopman operator $\clK$.
The main advantage of this approach is that the 
gain intensity can be easily 
obtained by the small time perturbation theory, 
\eqref{koop-16}, which yields 
\[\clA(t) \approx 
e^{4k^2}\frac{g^2t^2}{2} \sum_{j',j=1}^N
e^{2ik(\beta_j^*+\beta_j-\beta_{j'}^*+\beta_{j'})}.
\]
If initially, all coherent states are just the same, that is
$\beta_j=\beta, ~\forall j$, then according to
Eqs. \eqref{koop-8}, the distributions of the initial conditions
$p_j$ and $z_j$ are such that $(p_j^2+z_j^2)/2=\hbar|\beta|^2 ~~
\forall j$, which is defined $N^2$ points on the hyper-sphere 
in the $2N$ dimensional phase space. In other words, 
all $p_j$ and $z_j$ 
are independent initial values, defined on the 
hyper-sphere,
which is just the direct product of $N$ 
circles for the non-interacting electrons.
In this case the gained intensity is
\begin{equation}\label{disc-2}
\clA(t)\sim  g^2N^2t^2 ,
\end{equation}
which is \textit{superradiance}. 
Note that superradiance supposes that 
the intensity gain is of the order of $N^2$.

In the second approach, the evolution of the wave function is estimated in the short time limit, as well.
The obtained expression \eqref{fel-23} explains also 
the superradiance conditions, when the laser amplitude is 
the square of electron's number, $N^2$. 
Note also that the quantum mechanical analysis 
yields this superradiance condition \cite{BeMc83}. 
Let us estimate remained off-diagonal terms and consider the latter 
as follows
\begin{multline}\label{super-1}
g^2t^2\sum_{j=1}^{N}\sum_{l=1}^{N} \left(\prod_{s=1}^N\lgl p_{s}|\right)
e^{- 2k\hbar\frac{d}{dp_j}}
 e^{2ik\tau'\frac{\hatp_j}{m}-2ik\tau\frac{\hatp_l}{m}}  \\
\times e^{2k\hbar\frac{d}{dp_l}}
\left(\prod_{s=1}^N|p_s\rgl\right) ,
\end{multline}
where for different $s$ the wave functions are orthogonal, 
$\lgl p_s|p_{s'}\rgl=\delta_{s,s'}$ however $\lgl p_s|{p'}_s\rgl \neq 0$
for $p\neq p'$. Then Eq. \eqref{super-1} accounts all matrix 
elements including the off-diagonal elements as well. All 
the off-diagonal terms are either 
$
e^{it\frac{4k^2\hbar}{m}}
\lgl p_s+2k\hbar|p_s\rgl 
$
or
$
e^{-it\frac{4k^2\hbar}{m}}
\lgl p_s|p_s+2k\hbar\rgl , 
$
where $\lgl p_s+2k\hbar|p_s\rgl = \lgl p_s+2k\hbar|p_s\rgl =\mathcal{B}$.
Therefore, the laser amplitude reads
\begin{multline}\label{super-2}
\clA(t)_{\rm SR}=|\alpha|^2+ 2tg^2N
\\
+2g^2t^2\mathcal{B}^2N(N-1)
\cos\left[t\frac{4k^2\hbar}{m}\right]\\
\end{multline}

\section{Conclusion} 
A quantum model of a free-electron laser is considered 
for the many electron system described by Hamiltonian \eqref{fel-1}. 
A rigorous quantum mechanical analysis is performed for the calculation of the intensity laser field $\clA(t)$ 
in the basis of the coherent states both photon and electron.
Follow this formal result, we suggest a short time consideration.
This approximation is reasonable in the case when the speed of electrons is larger than the speed of light. In this case the gain 
mechanism of the FEL is the Cherenkov radiation.
In the framework of the suggested short time analysis 
the superradiance for the photon field intensity is obtained.

\appendix

\section{Quantum FEL Hamiltonian}\label{q-fel-H}

The basic Hamiltonian \eqref{fel-1}, which governs a variety of
different situations has been inferred in Refs. \cite{BeMc83,BeMc87}.
We also follow Ref. \cite{KGEPS15}.
The starting point is the classical relativistic Hamiltonian
\begin{equation}\label{q-fel-H-1}
H=[(c\mathbf{p} - e\mathbf{A})^2 + m_0^2 c^2]^{\frac{1}{2}},
\end{equation}
which describes a single electron with the rest mass $m_0$ and the elementary charge $e$ interacting with two electromagnetic plane
wave ﬁelds $\mathbf{A}=\mathbf{A}_L-e\mathbf{A}_W$. The laser field 
\begin{equation}\label{q-fel-H-2}
\mathbf{A}_L=\tilde{A}_L\mathbf{\ep}e^{-ik_L(ct - z)}+C.C. ,
\end{equation}
travels in positive $z$-direction, while the wigler fields
\begin{equation}\label{q-fel-H-3}
\mathbf{A}_W=\tilde{A}_W\mathbf{\ep}e^{-ik_W(ct + z)}+C.C. ,
\end{equation}
propagates in the opposite direction. Here 
$\tilde{A}_L,~\tilde{A}_W$ are amplitudes of the vector potentials,
$c$ is the speed of light, while the circular polarization 
is chosen, when $\mathbf{\ep}\cdot\mathbf{\ep}^*=1$ and 
$\mathbf{\ep}^2={\mathbf{\ep}^*}^2=0$.
In a so called Bambini–Renieri frame \cite{BaRe78,BaReSt79}, 
the wave numbers $k_ L$ and $k_W$ coincide, \textit{a.e},
$k_l=k_W=k$.

The non-relativistic approximation is performed by expanding the relativistic square root and neglecting the relativistic terms
that yields \cite{KGEPS15}
\begin{equation}\label{q-fel-H-4}
H ( z , p ) = \frac{p^2}{2m} +\frac{e^2}{m}\tilde{A}_w
\left(\tilde{A}_Le^{2ikz} + \tilde{A}_L^*e^{-2ikz}\right) ,
\end{equation}
where $m=m_0\sqrt{1+a_0^2}$  is the shifted mass
with the wiggler parameter \cite{KGEPS15,ScDoRo08}
\begin{equation}\label{q-fel-H-5}
a_0=\frac{\sqrt{2}e|\tilde{A}_W|}{m_0c} .
\end{equation}

\subsection{Quantization}
In the quantization procedure of the classical Hamiltonian 
\eqref{q-fel-H-5}, the electron dynamics and the laser field 
are quantized. Therefore the electron coordinate and the 
momentum are operators $(z,p)\rightarrow (\hat{z},\hat{p})$,
where their commutation relation is $[\hat{z},\hat{p}]=i\hbar$.
Then treating the laser ﬁeld amplitude $\tilde{A}_L$ as the 
quantized ﬁeld, the photon annihilation $\hat{a}$ and creation 
$\hat{a}^{dag}$ operators are introduced, 
and the commutation relation is $[\hat{a}, \hat{a}^{\dag}]= 1$.
Then the quantized amplitudes of the laser field are defined 
by the substitutions
\begin{align*}
&\tilde{A}_L\rightarrow A_L\hat{a} \\
&\tilde{A}_L^*\rightarrow A_L \hat{a}^{\dag} ,
\end{align*}
where $A_L$ is the amplitude of the quantized laser ﬁeld.
However, the wiggler field is considered 
as an external classical parameter $\tilde{A}_W^*=\tilde{A}_W=const$
\cite{KGEPS15}. 

Thus the quantized Hamiltonian reads
\begin{equation}\label{q-fel-H-6}
\hat{H}=\frac{\hat{p}^2}{2m}+\hbar g\left(\hat{a}e^{2ik\hat{z}}
+\hat{a}^{\dag}e^{-2ik\hat{z}} \right),
\end{equation}
where the coupling constant is
\begin{equation}\label{q-fel-H-7}
g=\frac{e^2}{\hbar m}A_L\tilde{A}_W ,
\end{equation}
which relates to the wiggler scenario of the FEL.
Eventually, straightforward generalization of the Hamiltonian 
\eqref{q-fel-H-6} for the $N$ electrons leads to the Hamiltonian 
\eqref{fel-1}. 

\subsection{Cherenkov radiation scenario of the FEL}

A slightly different consideration has been suggested in Ref. 
\cite{BeMc87}, which is also suitable for the Cherenkov 
radiation (CR)
scenario of the FEL. In the case of stimulated CR, 
the wiggler is replaced by a medium with a refractive index 
$n>1$. Since $\mathbf{A}_W=0$, the interaction term is due to the 
transverse canonical momentum of the electron 
$\mathbf{p}_{\bot}\neq 0$, while linear polarization 
$\mathbf{\ep}=\mathbf{\ep}^*$ is supposed
in Eq. \eqref{q-fel-H-2}. The relativistic Hamiltonian now reads
\begin{equation}\label{q-fel-H-8}
H=[c^2\mathbf{p}^2 - 2ce\mathbf{\ep}\mathbf{A}_L + 
m_0^2 c^4]^{\frac{1}{2}}.
\end{equation}
As shown in Ref. \cite{BeMc87}, quantization of the Hamiltonian
\eqref{q-fel-H-8} for the $N$ electrons systems yields
\begin{multline}\label{q-fel-H-9}
\hat{H}=\hbar\omega\hat{a}^{\dag}\hat{a} 
+\sum_j\frac{p_j^2}{2m} \\
+\hbar g\sum_j\left[\hat{a}e^{i(kz_j+\omega t)}
+\hat{a}^{\dag}e^{-i(kz_j+\omega t)}\right],
\end{multline}
where $\omega=ck/\sqrt{n^2-1}$ is the frequency of the laser field.
The coupling constant now is 
$g=-\frac{e}{\hbar mc}\mathbf{\ep}\cdot \mathbf{p}_{\bot}A_L$,
while $m^2=m_0^2+\mathbf{p}_{\bot}^2$,  \cite{BeMc87}.

In the interaction representation 
\[
\hat{H}\rightarrow e^{-i\omega\hat{a}^{\dag}\hat{a} t}\hat{H}
e^{i\omega\hat{a}^{\dag}\hat{a} t} 
\]
and rescaling $k\rightarrow 2k$, 
one arrives at the Hamiltonian $\hclH$ in Eq. \eqref{fel-1},
obtained by means of the commutation rule \cite{louisell},
\begin{equation}\label{q-fel-H-10}
e^{-\eta\hat{b}}\hat{c}e^{\eta\hat{b}}=
\sum_{n=0}^{\infty}\frac{(-\eta)^n}{n!}
[\overset{n}{\overbrace{\hat{b},[\hat{b}\dots[\hat{b}}},\hat{c}]\dots],
\end{equation}
where over-brace contains $n$ operators $\hat{b}$.
Taking into account that $\hat{b}\equiv \hat{a}^{\dag}\hat{a}$,
while $\hat{c}= (\hat{a}, \hat{a}^{\dag})$ 
\begin{align}\label{q-fel-H-11}
& e^{-i\omega\hat{a}^{\dag}\hat{a} t}\hat{a}
e^{i\omega\hat{a}^{\dag}\hat{a} t} =\hat{a}e^{-i\omega t} , \\
e^{-i\omega\hat{a}^{\dag}\hat{a} t}\hat{a}^{dag}
e^{i\omega\hat{a}^{\dag}\hat{a} t} =\hat{a}^{dag}e^{i\omega t} .
\end{align}

\section{Koopman operator}\label{koop-operator}

We present the inferring of $\clK_{e-ph}$ for a one electron system,
while its generalization for the $N$ electrons is straightforward.
The procedure is as follows. Performing first the average over the photon coherent states, where we define 
$\clA(\alpha)=\lAlpha\hat{\clA}(t)\rAlpha$, we have
\footnote{In Ref. \cite{iom2022} the Koopman operator is suggested without presenting the inferring procedure.}
\begin{multline}\label{koop12-1a}
\clK_{e-ph}\clA(t)=\frac{g}{i}\lgl\beta_j|
\left\{\alpha\left[\clA(\alpha) e^{2ik\hatz_j}-e^{2ik\hatz_j}\clA(\alpha)\right] \right. \\
+ \left.\alpha^*\left[\clA(\alpha)e^{-2ik\hatz_j}-e^{-2ik\hatz_j}\clA(\alpha)\right] \right. \\
+\left.\left[\partial_{\alpha}\clA(\alpha)e^{-2ik\hatz_j}-
e^{2ik\hatz_j}\partial_{\alpha^*}\clA(\alpha)\right]\right\}
|\beta_j\rgl .
\end{multline}
The last line is the part of the Koopman operator
in Eq. \eqref{koop-12b}. The terms with $\alpha$ and $\alpha^*$
vanish. Let us show this. Note that
$D(\pm i\bar{k})=e^{\pm 2ik\hatz_j}=e^{\pm i\bar{k}(\hb_j+\hbdag_j)}$ with $\bar{k}=2k\sqrt{\hbar/2}$
is a shift operator for the coherent states $|\beta\rgl$ 
\cite{perelomov}. 
Namely 
\begin{equation}\label{koop-12-2a}
e^{\pm 2ik\hatz_j}|\beta_j\rgl=e^{i\mIm(\pm i\bar{k}\beta_j^*)}
|\beta+(\pm i\bar{k})\rgl.
\end{equation}
Therefore, we have
\begin{subequations}\label{koop-12-3a}
\begin{align}
&\lgl\beta_j|\clA(\alpha) e^{\pm 2ik\hatz_j}|\beta\rgl 
= e^{i\mIm(\pm i\bar{k}\beta_j^*)}\lgl\beta_j|\clA(\alpha)
|\beta+(\pm i\bar{k})\rgl \nonumber  \\
&= e^{i\mIm(\pm i\bar{k}\beta_j^*)}
\clA(\alpha, \beta^*,\beta \pm i\bar{k})
\lgl\beta_j|\beta+(\pm i\bar{k})\rgl , 
\label{koop-12-3a-1} \\
 &\lgl\beta_j|e^{\pm 2ik\hatz_j}\clA(\alpha)|\beta\rgl
=\clA(\alpha,\beta^*\mp i\bar{k})e^{\pm i\mIm(i\bar{k}\beta_j)} 
\nonumber \\
&\times\lgl\beta_j\mp(i\bar{k}|\beta)\rgl
\label{koop-12-3a-2}
\end{align}
\end{subequations}
Then demanding that expressions in Eq. \eqref{koop-12-3a-1}
equal to Eq. \eqref{koop-12-3a-2}, 
namely 
\begin{multline}
e^{i\mIm(\pm i\bar{k}\beta_j^*)}
\clA(\alpha, \beta^*,\beta \pm i\bar{k})
\lgl\beta_j|\beta\pm i\bar{k}\rgl \\
=
\clA(\alpha,\beta^*\mp i\bar{k})e^{i\mIm(\pm i\bar{k}\beta_j)}
\lgl\beta_j-(\pm i\bar{k})|\beta\rgl,
\end{multline}
which immediately follows from the real mean physical values, 
we arrive at the Koopman operator \eqref{koop-12}.



\bibliography{bib-Koopman}


\end{document}